\documentclass[]{spie}  

\usepackage{amsmath,amsfonts,amssymb}
\usepackage{graphicx}
\usepackage{url}
\usepackage[colorlinks=true, allcolors=blue]{hyperref}

\title{New techniques for high-resolution imaging and high-precision wavefront sensing via masked-aperture interferometry}

\author[a]{Nithyanandan Thyagarajan}
\author[b]{Bojan Nikolic}
\author[c]{Chris Carilli}
\author[d]{Laura Torino}
\author[d]{Ubaldo Iriso}
\affil[a]{Space \& Astronomy, Commonwealth Scientific and Industrial Research Organisation (CSIRO), P. O. Box 1130, Bentley, WA 6102, Australia}
\affil[b]{Astrophysics Group, Cavendish Laboratory, University ofCambridge, Cambridge CB3 0HE,UK}
\affil[c]{National Radio AstronomyObservatory, P.O. Box O, 1011 Lopezville Road, Socorro, New Mexico 87801, USA}
\affil[d]{ALBA-CELLS Synchrotron Radiation Facility, Carrerde la Llum 2-26, Cerdanyola del Vall\`{e}s, 08290 Barcelona, Spain}

\authorinfo{Further author information: (Send correspondence to N.T.)\\N.T.: E-mail: Nithyanandan.Thyagarajan@csiro.au, Telephone: +61 8 6436 8626}

\begin{document} 
\maketitle

\begin{abstract}
Achieving high–angular-resolution imaging on subarcsecond scales is fundamentally limited by wavefront aberrations imparted by the propagation medium and by optical elements along the light path. Accurate recovery of source structure at these fine angular scales therefore relies on precise, real-time sensing and correction of wavefront aberrations. Drawing inspiration from radio interferometry, we have developed different approaches using masked apertures that directly measure both the amplitude and phase of the distortions to the electromagnetic wavefront while simultaneously reconstructing the underlying source structure. First is radio interferometry style self-calibration which can recover the complex electric field aberrations across the aperture with subarcsecond phase accuracy, equivalent to nanometer-level precision in optical pathlength, and simultaneously reconstruct the source structure with milliarcsecond (sub-micron) accuracy. Second is closure invariant-based source reconstruction which allows to bypass self-calibration and errors therein entirely while achieving comparable fidelity in the recovered source structure. These methods have been validated on the ALBA synchrotron beamline. Together, these methods provide a reliable framework for nanometre-scale high-precision wavefront sensing and subarcsecond-scale high angular resolution imaging, enabling new possibilities for masked-aperture interferometry. Potential applications span laboratory and synchrotron facilities to astronomy, including beam diagnostics in the Large Hadron Collider and masked-aperture interferometry on space telescopes such as the James Webb Space Telescope.

\end{abstract}


\section{INTRODUCTION}
\label{sec:intro}  

Optical and radio interferometry share a common theoretical foundation in Fourier optics and the van Cittert–Zernike theorem \cite{Zernike_1938}, which establishes the relationship between the mutual coherence function measured across an aperture and the brightness distribution of the source. Despite this common basis, the two fields have evolved largely independently because of fundamental differences in how electromagnetic radiation is detected and processed.

Recent developments in computational imaging and optical interferometry have begun to narrow this gap by borrowing concepts from radio astronomy. The mathematical equivalence of the underlying interferometric measurement equations suggests that many of the calibration and imaging techniques developed for radio arrays can be adapted to optical systems, provided appropriate measurement strategies are employed.

In this work, we adapt several powerful radio interferometric techniques to masked-aperture optical interferometry. Specifically, we investigate radio-style self-calibration for direct estimation of complex wavefront aberrations, the use of calibration-independent closure invariants for robust image reconstruction, and a generalized self-calibration framework applicable to both non-redundant and redundant aperture masks. These approaches enable simultaneous high-precision wavefront sensing and high-angular-resolution imaging, providing improved characterization of optical beams in synchrotron facilities while offering a pathway toward applications in astronomical instrumentation, including sparse-aperture interferometry and wavefront sensing for future space- and ground-based telescopes.

The paper is organised as follows. Section~\ref{sec:interferometry} introduces interferometry contrasting its application in radio and optical wavelengths. Section~\ref{sec:methods} describes the methods we have adapted from radio interferometry to aperture masking at optical wavelengths. The results and ongoing work are described in section~\ref{sec:results}. We summarise the work in section~\ref{sec:summary}.

\section{Interferometry in radio and optical wavelengths}\label{sec:interferometry}

The van Cittert-Zernike theorem \cite{Zernike_1938} states that the spatial coherence of the complex electric fields in the aperture plane is related to the intensity distribution in the image plane, which under certain approximations reduces to a Fourier transform relationship \cite{TMS2017}. The relationship can be mathematically expressed as
\begin{align}
    V_{ab}(\lambda) &\equiv \langle E_a(\lambda) E_b^*(\lambda)\rangle = \int_\Omega P_{ab}(\hat{\boldsymbol{s}},\lambda) \, I(\hat{\boldsymbol{s}},\lambda) e^{-i 2\pi \boldsymbol{u}_{ab}\cdot \hat{\boldsymbol{s}}} \, \mathrm{d}\Omega\, , \label{eqn:VCZ}
\end{align}
where, $I(\hat{\boldsymbol{s}},\lambda)$ denotes the intensity distribution on source plane, $\Omega$, at location $\hat{\boldsymbol{s}}$ and wavelength, $\lambda$. $a$ and $b$ denote two elements in the aperture, whose effective directional sensitivity is given by $P_{ab}(\hat{\boldsymbol{s}},\lambda)$. $E_a(\lambda)$ denotes the electric field if it could be sampled at the aperture element $a$. Finally, $V_{ab}(\lambda) \equiv \langle E_a(\lambda) E_b^*(\lambda)\rangle$ denotes the spatial coherence of the electric fields at aperture elements $a$ and $b$, and is also referred to as \textit{visibility}. It is evident that Equation~\ref{eqn:VCZ} closely resembles a spatial Fourier transform.

Optical interferometry, in most cases, measures only the intensity of the electromagnetic field in the image or focal plane because optical frequencies are too high for direct electronic measurement of the electric field. Recovering phase information,  therefore, requires indirect approaches based on interference, phase retrieval, or wavefront sensing. While remarkable advances have been made in adaptive optics, aperture masking, and optical interferometry, calibration of wavefront distortions and recovery of high-fidelity source structure remain challenging, particularly in photon-limited observations or when optical aberrations vary rapidly.

Contrastingly, in radio interferometry \cite{TMS2017}, the electric field represented by complex voltages, consisting of both amplitude and phase, from individual antennas sampling the aperture can be coherently amplified, digitised, and cross-correlated to estimate the complex visibility function in the aperture (Fourier) plane. This capability has enabled the development of sophisticated calibration and imaging techniques that jointly estimate instrumental and propagation-induced corruptions while reconstructing the source brightness distribution. In particular, self-calibration \cite{TMS2017}, closure invariants \cite{Thyagarajan+2022,Samuel+2022} like closure phases \cite{Jennison1958,Thyagarajan_Carilli2022}, and closure amplitudes \cite{Twiss+1960} have become indispensable tools for high-fidelity imaging, allowing radio interferometers to produce diffraction-limited images even in the presence of significant instrumental and atmospheric errors.

\section{Application: Synchrotron from particle accelerators}\label{sec:application}

In this paper, we describe recent efforts that have adapted radio interferometric techniques to the optical interferometry context found specifically in particle accelerator facilities used in the production of synchroton radiation. Accurate characterization of charged particle beams is a fundamental diagnostic requirement in modern particle accelerators and synchrotron light sources \cite{Wiedemann2015}. The transverse beam profile directly influences the quality, brightness, and stability of the emitted synchrotron radiation, making precise beam monitoring essential for accelerator tuning, beam optimisation, and reliable facility operation. Beam dimensions must be controlled at the micrometre scale, often requiring optical wavefront aberrations to be measured and corrected with nanometre-level precision to ensure efficient beam transport and delivery. Smaller and more stable beams produce tighter optical focusing, higher photon flux densities, and improved experimental performance across a broad range of scientific applications. A particularly valuable approach for beam diagnostics is synchrotron radiation interferometry (SRI), which provides a non-invasive means of characterising the electron beam by analysing the spatial coherence of the emitted synchrotron radiation \cite{Kube2007}. 

Until recently, common SRI implementations employed a dual-mask aperture, similar to Young's double-slit interferometer. The measured fringe visibility provides information about the beam size along the direction perpendicular to the slit separation through the van Cittert–Zernike theorem \cite{Zernike_1938}. Two-dimensional beam profiles are typically obtained by mechanically rotating the aperture or repeating measurements with different slit orientations \cite{Torino+Iriso2016}. This approach is inherently sequential, primarily because of the use of a two-hole aperture, which in turn is necessitated because of wavefront distortions and loss of coherence that are inherent when using a multi-mask aperture. Thus, such approaches provide only indirect reconstruction of the transverse beam distribution, and offers limited capability for characterising optical aberrations introduced by the beamline.

These limitations motivated the development of a more general interferometric framework capable of simultaneously reconstructing the two-dimensional beam profile while estimating the wavefront distortions across the optical aperture. Such a capability would enable instantaneous beam characterisation, eliminate the need for mechanical aperture rotation, and provide direct information about both the source structure and the optical system.

Radio interferometry has long addressed analogous challenges in the presence of not just two but even hundreds of aperture elements through sophisticated calibration and imaging techniques, including self-calibration and closure invariants, which enable high-fidelity image reconstruction despite significant instrumental and propagation-induced errors \cite{TMS2017}. The mathematical basis shared between radio and optical interferometry suggests that these methods can be translated to synchrotron radiation interferometry to provide robust, calibration-based as well as calibration-immune beam diagnostics.

The objective of this work is therefore to develop and experimentally demonstrate a unified interferometric framework that adapts these radio interferometric techniques to optical SRI. Our approach simultaneously reconstructs the transverse beam profile while estimating the complex amplitude and phase distribution across the aperture, enabling robust beam characterisation and precision wavefront sensing from a single interferometric measurement.

\section{Methods}\label{sec:methods}

\begin{figure} [ht]
   \begin{center}
   \begin{tabular}{cc} 
   \includegraphics[height=5cm]{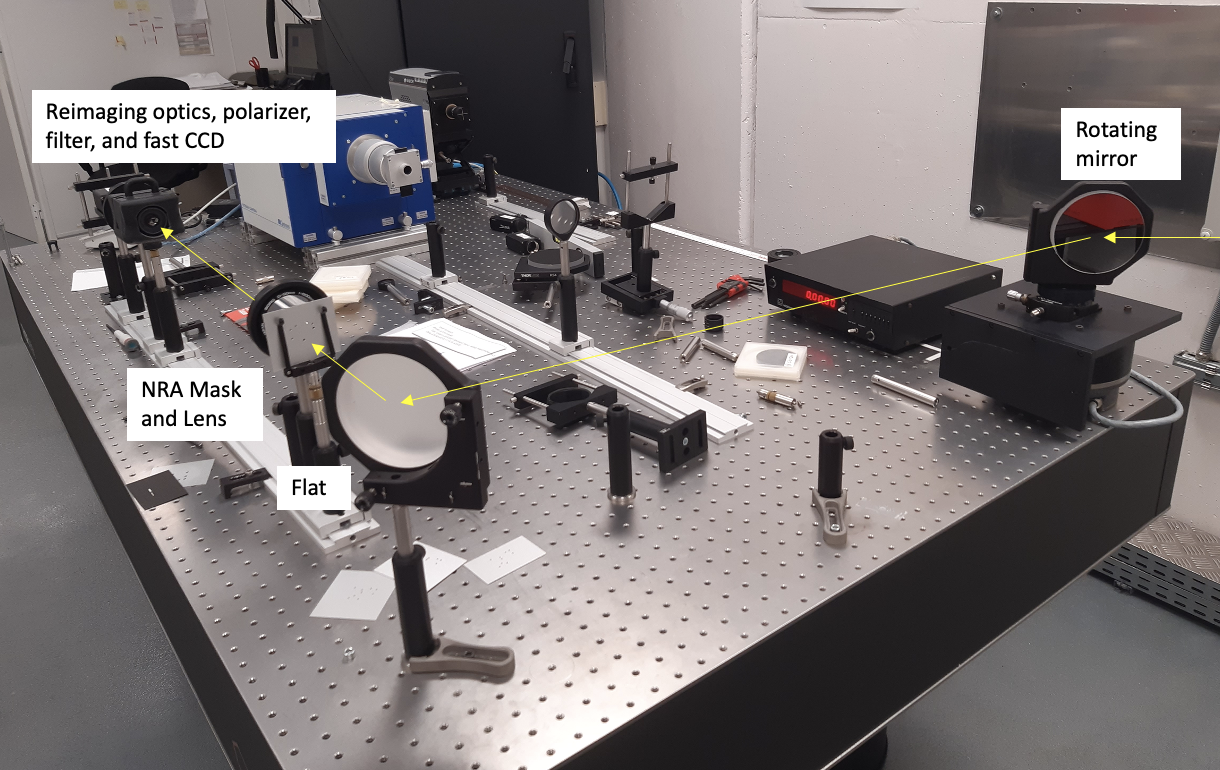} & 
   \includegraphics[height=3.5cm]{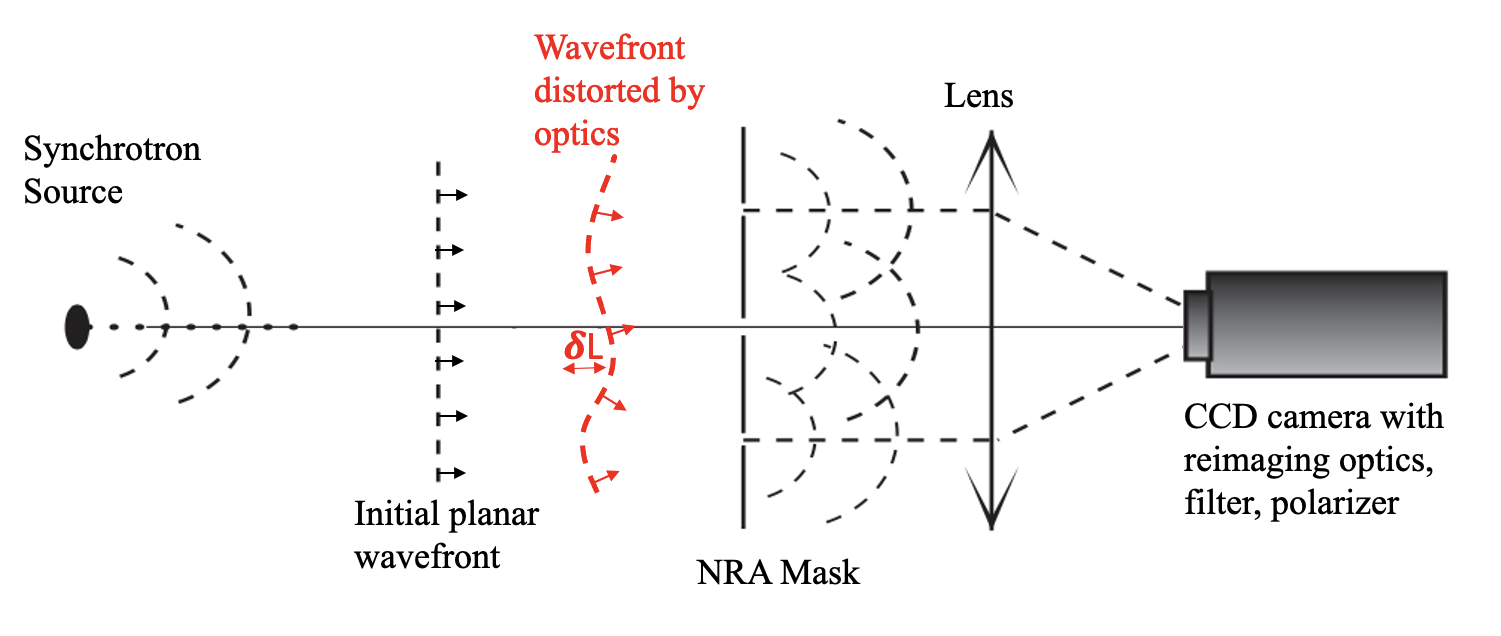}  
   \end{tabular}
   \end{center}
   \caption[Experimental setup] 
   {\label{fig:setup} Experimental setup at the Xanadu optical bench at the ALBA synchrotron facility (reproduced with permission from Carilli et al. (2025)\cite{Carilli+2025}).}
\end{figure}

The experimental setup used in the Xanadu optical bench at the ALBA synchrotron facility is illustrated in Figure~\ref{fig:setup}. 
At the aperture, we employ non-redundant aperture (NRA) masks, in which every pair of aperture masks forms a unique baseline. Consequently, each interference fringe corresponds to a distinct spatial frequency in the Fourier plane, ensuring that the measured Fourier components are uniquely associated with individual aperture pairs. 
Unlike redundant aperture configurations, where multiple aperture pairs may or may not contribute coherently to the same Fourier mode and complicate the interpretation of the measured interferogram, non-redundant masks provide a direct and unambiguous mapping between the measured visibilities and the sampling locations in the Fourier plane. In this work, five-hole and seven-hole NRA masks were designed to simultaneously sample multiple spatial frequencies while maintaining this one-to-one correspondence. 

This unique mapping allows the interferometric measurements to be interpreted in exactly the same way as the pairwise visibility measurements obtained by a radio interferometer. Each aperture samples the incident electromagnetic wavefront, and the interference between every pair of apertures yields a complex visibility that represents one Fourier component of the source brightness distribution. However, these measured visibilities are corrupted by variations in the complex electric field across the aperture arising from non-uniform illumination and wavefront aberrations. 

This relationship can be expressed in the same form as the radio interferometric measurement equation,
\begin{align}
    V_{ab}^\textrm{obs} &= G_a G_b^* V_{ab} \, , \label{eqn:gain-corruptions}
\end{align}
where, $V_{ab}^\textrm{obs}$ denotes the measured visibility between aperture elements $a$ and $b$, $V_{ab}$ is the true source visibility, and $G_a$ is the complex gain associated with aperture element $a$. The complex gain encapsulates both the amplitude illumination and the phase distortion of the wavefront at that aperture location. While these complex gains are traditionally interpreted as instrumental and atmospheric corruptions in radio astronomy, they do not occur naturally in optical measurements because photon phases are not measured. However, the squared amplitudes of the gains, $|G_a|^2$, naturally represent illumination at the aperture masks and the gain phases represent optical aberrations in masked-aperture interferometry.

This formulation enables direct adaptation of radio interferometric calibration techniques to optical interferometry. In recent works, we designed both five-hole and seven-hole non-redundant aperture (NRA) masks to sample multiple spatial frequencies simultaneously. These measurements are used to reconstruct the transverse beam profile while simultaneously estimating the complex aperture gains. In addition, calibration-independent closure quantities are employed both to validate the recovered solutions and to reconstruct the source without explicit gain estimation.

\begin{figure} [ht]
   \begin{center}
   \begin{tabular}{c} 
   \includegraphics[height=12cm]{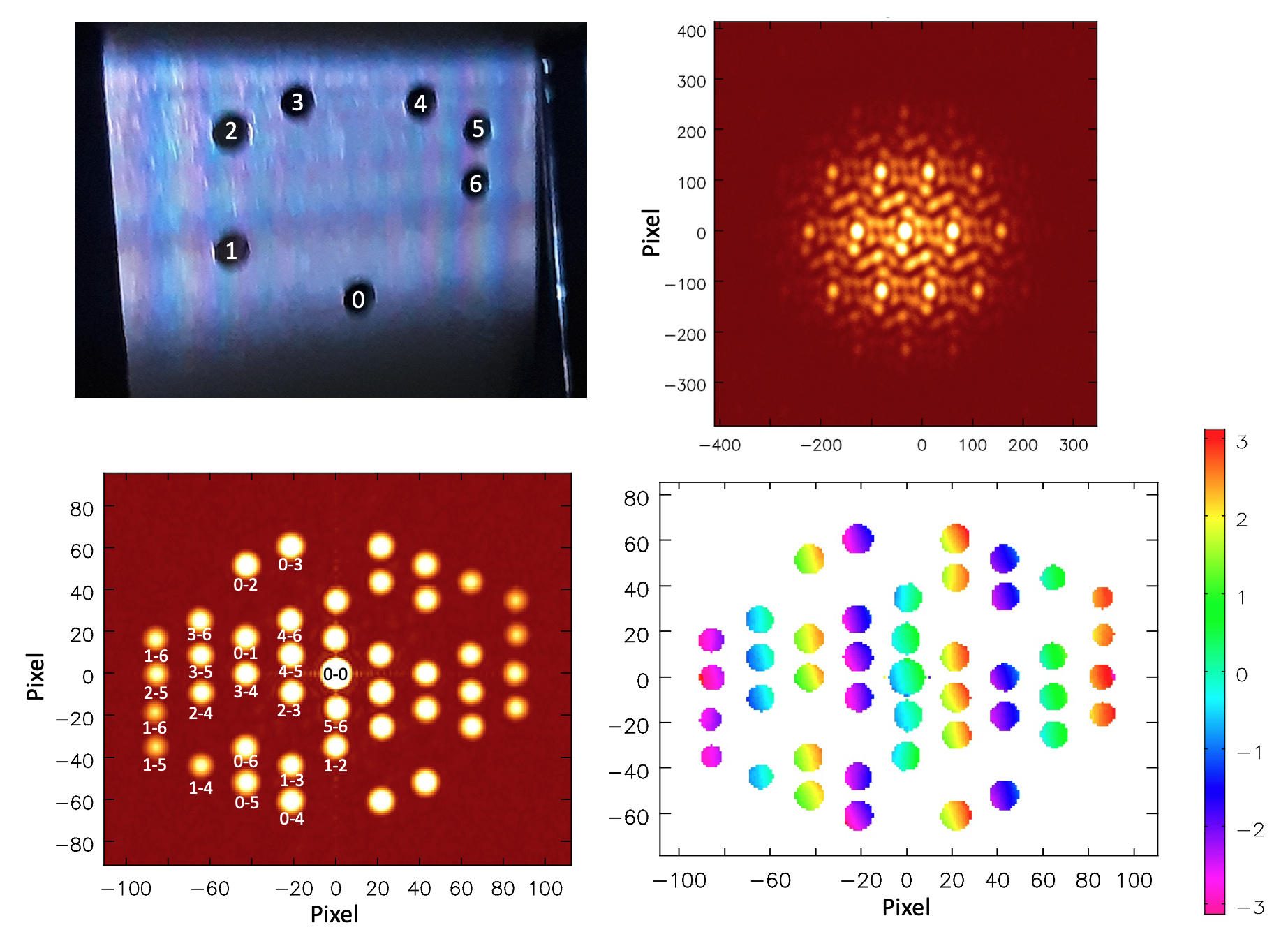} 
   \end{tabular}
   \end{center}
   \caption[Image and Fourier plane views] 
   {\label{fig:measurements} \textit{(Top left):} A 7-hole non-redundant aperture mask. A non-uniform illumination across the aperture is visible signifying amplitude corruptions on the incident wavefront. Phase corruptions are also present but not shown. \textit{(Top right):} One frame of interferogram image recorded on the CCD. \textit{(Bottom left):} Amplitude of the Fourier transform in the Fourier plane of the interferogram. Each distinct blob is a visibility corresponding to a unique hole pair as annotated. The central blob denoted by ``0-0'' represents the sum of intensities through all the holes. \textit{(Bottom right):} Phase of the Fourier transform in the Fourier plane of the interferogram. (reproduced with permission from Carilli et al. (2025)\cite{Carilli+2025}).}
\end{figure}

Figure~\ref{fig:measurements} shows the complex visibilities measured using the NRA mask on the Fourier plane using a Fourier transform of the interferogram images recorded on the CCD. The wavelength of light was 538~nm. The top left panel in Figure~\ref{fig:measurements} illustrates the illumination pattern across the aperture that is represented by $|G_a|^2$ at the mask locations. The phase aberrations across the aperture, while certainly present, are not shown here. 

\subsection{Self-calibration for simultaneous beam reconstruction and wavefront sensing}\label{sec:self-cal}

Our first approach adapts the self-calibration framework developed for radio interferometric imaging. Rather than treating the aperture gains as nuisance parameters, the algorithm jointly estimates the source brightness distribution together with the complex gain associated with every aperture element by minimising 
\begin{align}
    \chi^2 &= \sum_a\sum_b |G_a G_b^* V_{ab} - V_{ab}^\textrm{m}|^2 \label{eqn:optimisation}
\end{align}
with respect to $G_a$ and the source model parameters in $V_{ab}^\textrm{m}$. Traditionally, the optimisation has alternated between estimating the source structure from the current gain estimate and updating the complex gains using the reconstructed source model. This iterative process converges to a mutually consistent solution that simultaneously explains the measured interferograms and the observed fringe visibilities. More recent optimisation alternatives offer simultaneous estimation of the gain and source model parameters, particularly when the number of measurements exceed the number of estimated parameters. 

The recovered complex gains provide direct physical information about the optical system. Their squared amplitudes represent the illumination pattern across the aperture, while their phases correspond to the optical wavefront distortion sampled at each aperture location. Consequently, a single self-calibration procedure yields the two-dimensional transverse beam intensity distribution, the aperture illumination function, and the wavefront phase distribution, thereby combining beam diagnostics and precision wavefront sensing within a unified reconstruction framework \cite{Carilli+2025,Carilli+2026}.

If the beam has a Gaussian cross-section, as is the case with the ALBA synchrotron beam, the gain phases do not constrain the Gaussian source parameters and only the amplitudes of the gains (one for each hole in the mask) and the Gaussian source parameters (major and minor axes, and orientation angle of the ellipse) need to be estimated in the optimisation process \cite{Nikolic+2024}, which reduces to \textit{amplitude self-calibration}. However, knowledge of wavefront errors can still be obtained from estimating the gain phases. 

\subsection{Source reconstruction from calibration-independent closure invariants}\label{sec:ClAmp}

Although self-calibration estimates the complex aperture gains directly, radio interferometry has long recognised that certain combinations of visibilities are inherently independent of these gains. These calibration-independent observables, known as closure invariants \cite{Thyagarajan+2022,Samuel+2022}, provide robust constraints on the source structure even when the individual aperture gains are unknown. The popular variants of closure invariants have been known as closure phases \cite{Jennison1958,Thyagarajan_Carilli2022} and closure amplitudes \cite{Twiss+1960}.

In recent work \cite{Thyagarajan+2025}, we employed closure amplitudes formed from products and ratios of measured visibility amplitudes,
\begin{align}
    A_{mnpq}^\textrm{obs} &= \frac{|V_{mn}^\textrm{obs}| \, |V_{pq}^\textrm{obs}|}{|V_{mp}^\textrm{obs}| \, |V_{nq}^\textrm{obs}|}
\end{align}
Because the complex aperture gains cancel algebraically, these quantities depend only on the intrinsic source structure and are, therefore, insensitive to variations in aperture illumination or wavefront errors. Only closure amplitudes were employed because the synchrotron beam cross-section was known to be a Gaussian. For reconstructing more complex source structures, the above process can be generalised to include a combination of closure phases and closure amplitudes. 

The use of closure amplitudes serves two complementary purposes. First, they provide an independent reconstruction pathway that does not require self-calibration, allowing the beam profile to be recovered directly from calibration-independent observables. Second, they provide an important validation of the amplitude self-calibration results by verifying that the recovered beam structure is consistent with observables that are immune to gain errors.

\section{Results}\label{sec:results}

The proposed framework was evaluated using interferometric measurements acquired with both five-hole and seven-hole non-redundant aperture (NRA) masks on the ALBA synchrotron beamline. In both configurations, the unique Fourier-plane sampling provided by the NRA masks enabled simultaneous measurement of multiple spatial frequencies from a single interferogram, allowing instantaneous two-dimensional characterization of the synchrotron beam.

The self-calibration framework successfully reconstructed the transverse beam intensity distribution while simultaneously estimating the complex aperture gains. The recovered beam profiles agreed closely with independent reference measurements, achieving an accuracy of approximately 1\%, corresponding to a spatial accuracy of roughly 1~$\mu$m in the reconstructed beam dimensions, that corresponds to subarcsecond accuracy in angular dimensions. In addition to recovering the beam structure, the estimated gain amplitudes provided the incident illumination pattern across the aperture.

\begin{figure} [ht]
   \begin{center}
   \begin{tabular}{cc} 
   \includegraphics[height=5cm]{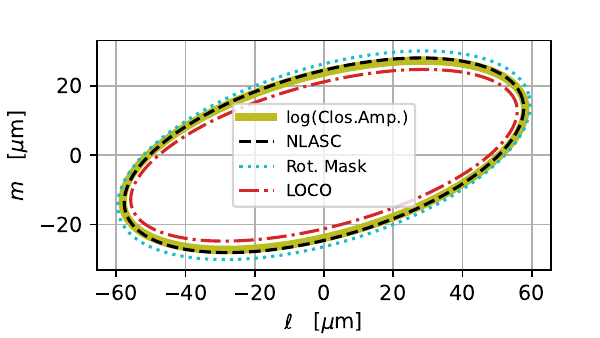} & 
   \includegraphics[height=5cm]{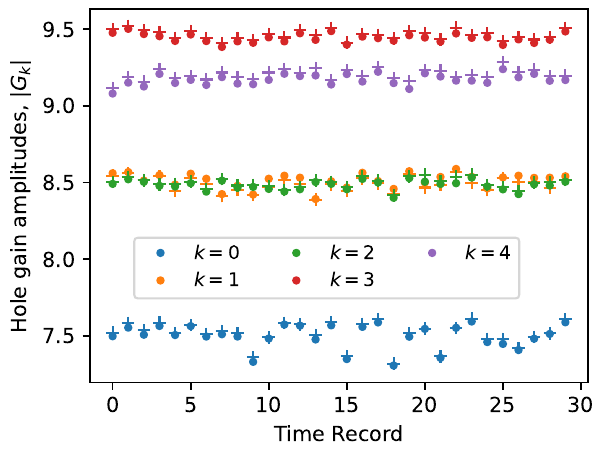}
   \end{tabular}
   \end{center}
   \caption[Two-dimensional beam cross-section] 
   {\label{fig:gaussian-beam} \textit{(Left):} Independent estimates of Gaussian synchrotron beams with the amplitude self-calibration \cite{Nikolic+2024} (``NLASC'' in legend) and closure amplitude \cite{Thyagarajan+2025} (``log(Clos.Amp.)'') methods. Also included are other standard methods for comparison, namely, the rotated two-hole mask \cite{Torino+Iriso2016} (``Rot. Mask'' in legend) and beam simulation (``LOCO'') not part of this study. \textit{(Right:)}  Amplitudes of the gains at the five holes for 30 frames determined  using linear (plus symbols) and logarithmic (diamond symbols) variants of the amplitude self-calibration method \cite{Nikolic+2024,Thyagarajan+2025} (reproduced with permission from Thyagarajan et al. (2025)\cite{Thyagarajan+2025}).}
\end{figure}

An independent reconstruction using calibration-independent closure amplitudes produced beam profiles consistent with those obtained from self-calibration, demonstrating comparable reconstruction accuracy. The agreement between these two fundamentally different approaches validates both the robustness of the self-calibration algorithm and the utility of closure invariants for optical interferometry. The above results are illustrated in Figure~\ref{fig:gaussian-beam}. 

\begin{figure} [ht]
   \begin{center}
   \begin{tabular}{c} 
   \includegraphics[height=10cm]{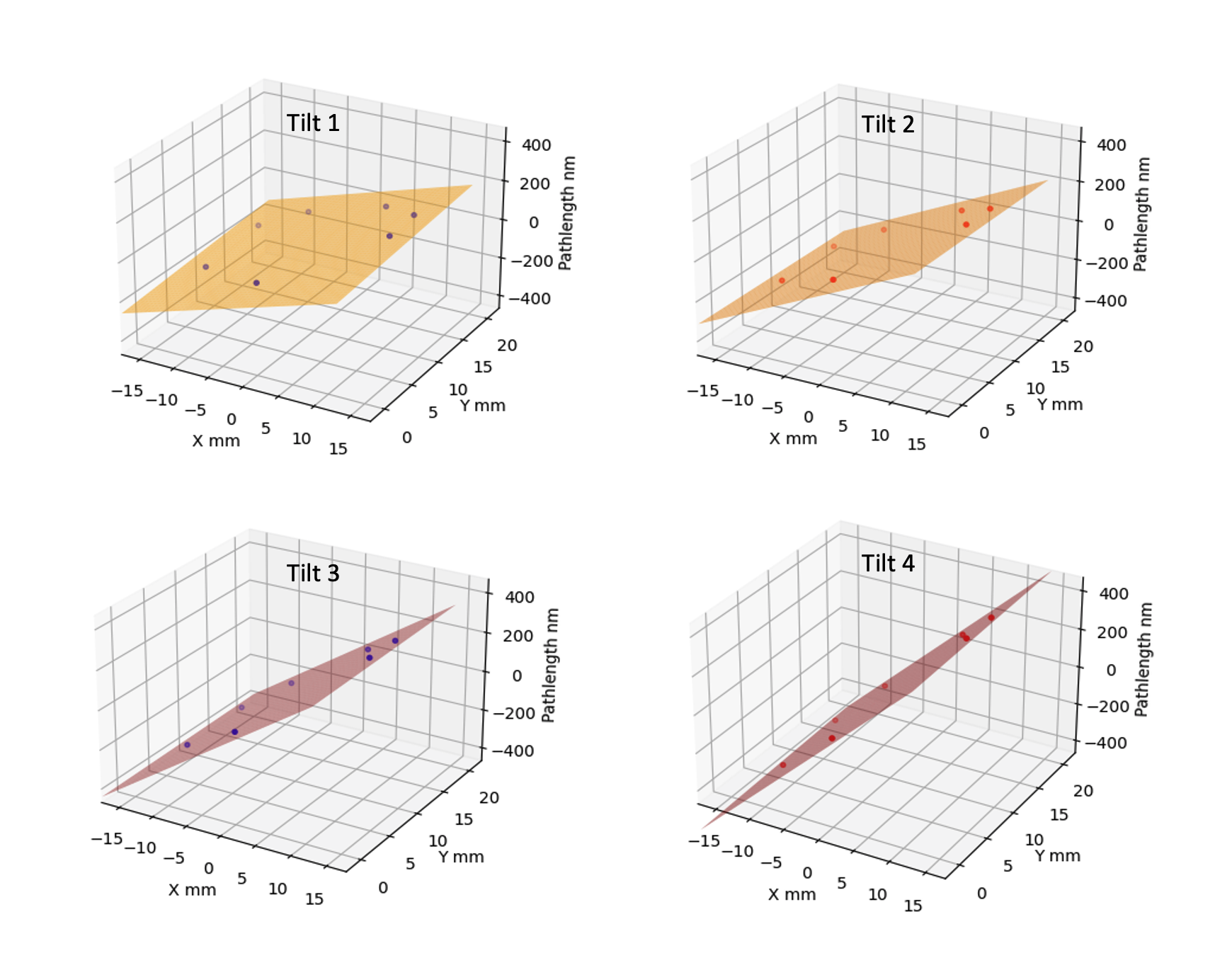} 
   \end{tabular}
   \end{center}
   \caption[Phases at the aperture] 
   {\label{fig:wfs} Tip-tilt pathlengths of the photon wavefront through the optical system for four different mirror rotations ($1''$, $2''$, $3''$, and $4''$), as derived from the self-calibration gain phases and converted to pathlengths from a 7-hole mask aperture measurement. The wavefront deviations at the hole locations are well fit by a plane as expected for a tip-tilt caused by the mirror rotation (reproduced with permission from Carilli et al. (2025)\cite{Carilli+2025}).}
\end{figure}

To estimate the recovery of wavefront errors in the 7-hole measurements, a rotating flat mirror was used in the photon path whose angle of rotation can be controlled at set increments. This allowed us to introduce known levels of tip-tilt wavefront errors in the measurements. In this case, we used light at 400~nm. The recovered complex gain phases from the self-calibration procedure correspond directly to the wavefront distortions sampled at each aperture location. These phase estimates enabled reconstruction of the optical wavefront with nanometre-scale optical path-length precision, providing information on both wavefront aberrations and surface errors within the optical system. The results of wavefront sensing from the estimated gain phases are shown in Figure~\ref{fig:wfs}. Although a tip-tilt effect is well fit by a plane as demonstrated, the procedure is applicable without any loss of generality to more complex wavefront shapes. Further work of fitting more complex wavefronts using Zernike polynomial basis is underway. 


The simultaneous recovery of aperture illumination, wavefront phase, and beam structure demonstrates that the complex gains represent physically meaningful properties of the optical system. The agreement between the closure-based and self-calibrated reconstructions demonstrates both the robustness of the proposed framework and the applicability of radio interferometric techniques to optical masked-aperture interferometry.

Overall, these results demonstrate that radio interferometric calibration and closure techniques can be successfully translated to masked-aperture optical interferometry. A single interferometric measurement provides simultaneous high-accuracy beam diagnostics and precision wavefront sensing, eliminating the need for sequential measurements or independent wavefront metrology while providing a robust framework for radio-style calibration of the effects of the optics and interferometric imaging at optical wavelengths.

\section{Summary}\label{sec:summary}

We have demonstrated that calibration and imaging techniques developed for radio interferometry can be successfully translated to high signal-to-noise optical interferometry through masked-aperture measurements. By exploiting the mathematical equivalence between radio and optical interferometric observables, self-calibration and closure-based methods can be adapted to simultaneously recover source structure and estimate the complex wavefront across the aperture.

The use of non-redundant aperture masks enables unique sampling of the Fourier plane, allowing instantaneous reconstruction of the two-dimensional beam profile while simultaneously recovering the aperture illumination pattern and optical wavefront aberrations. Self-calibration provides a unified framework for estimating both the beam structure and the complex aperture gains, whereas closure amplitudes and closure phases offer calibration-independent observables that enable robust image reconstruction and independent validation. Together, these techniques achieve micron-scale high-resolution beam imaging and wavefront sensing with nanometre-scale optical pathlength precision from a single interferometric measurement.

The proposed framework is entirely non-invasive and is well suited to near real-time beam diagnostics, making it attractive for next-generation accelerator facilities where precise control of beam quality is essential. Beyond synchrotron radiation interferometry, the methodology is readily generalisable to other optical sources, including plasma-based light sources, X-ray free-electron lasers (XFELs), and optical interferometry on the James Webb Space Telescope (JWST), where simultaneous source characterization and wavefront sensing are of growing importance.

The techniques are currently being evaluated for beam diagnostics at the Large Hadron Collider (LHC), including applications to proton and heavy-ion beams where accurate characterization of beam halo and core structure is critical \cite{Torino+2025}. Looking ahead, future work will extend the framework to redundant aperture designs, which offer substantially higher photon throughput. This will require joint estimation of beam structure and wavefront errors in the presence of redundant Fourier measurements, bringing the full power of radio interferometric calibration techniques to more photon-limited optical interferometric systems.

 
\acknowledgments 
The National Radio Astronomy Observatory is a facility of the National Science Foundation operated under cooperative agreement by Associated Universities, Inc.
Related patents and patent applications: Patent No. US 12,104,901 B2; Provisional Patent No. 63/355,174 (US Patent application No. 18/878,488); Provisional Patent No. 63/648,303 (US Patent application No. 19/079,876).

\bibliography{refs} 
\bibliographystyle{spiebib} 

\end{document}